# Spatial pattern and power function deviation in a cellular automata model of an ant population

**John Vandermeer**
*Ecology and Evolutionary Biology, and Program in the Environment, University of Michigan, Ann Arbor, MI, 48109*

**and**

**Doug Jackson**
*Eastern Research Group, Inc., 110 Hartwell Avenue, Lexington, MA 02421-3136*

**Abstract**
*A spatially explicit model previously developed to apply to a population of arboreal nesting ants in southern Mexico is explored in detail, including notions of criticality and robust criticality and expected power law adjustment.*

**Introduction**
Spatial pattern has long been a preoccupation of ecologists, from the observational work of early botanists (e.g. Clements, 1916; Shelford, 1913) to the theoretical explorations of modern theoreticians (Tilman and Karieva, 1997; Levin, 1992; Dieckmann et al., 2000). In this note we elaborate on the details of a stochastic cellular automata model previously developed (Vandermeer et al., 2008) and thought to adequately represent the spatial dynamics of a population of a tropical ant species.

Non-random spatial patterns are ubiquitous in nature. Some are evidently a consequence of exogenous forcing from environmental variables such as moisture, temperature, substrate, and so forth. Others seem to emerge as a consequence of the dynamic interaction of the organisms of concern, of special interest when interactions occur at a local level and pattern emerges at a larger scale, frequently referred to as self-organization. Self-organization in spatial ecology has been reported for both spatial models (e.g., Pascual et al. 2002; Kefi et al., 2011; Rietkerk et al., 2004; Roy et al., 2003) and field data (Vandermeer et al., 2008; Wooten, 2001). One framework that seems to correspond qualitatively with many natural circumstances is an arrangement that mimics the so-called Turing mechanism in which foci expand locally (metaphorically acting like Turing's activators) but are restricted as the clusters formed from those foci become subject to some controlling force (metaphorically acting like Turing's repressor). An example from classical ecological thought is the Janzen/Connell effect, in which plants produce seeds that, when falling near the parent, produce a locally abundant seedling population which in turn is attacked by a predator or disease of the seedling (or seed) (Martin and Canham, 2010; Peterman et al., 2008). Plant dispersal thus is the activating process and seed or seedling predation is the repressive force. Such a dynamic, when applied on a lattice, eventually produces a distinctly non-random distribution of cluster sizes, organized only from the local dynamics of the system, which is to say, self-organized. Under certain conditions, that self-organization approaches a critical state in which the distribution of cluster sizes is similar at all scales, thus following a power law. In this latter case, the self-organization is referred to as self-organized criticality.

In classical formulations of criticality, a scale-free distribution is generated at some set of parameter values, and it is normally the case that deviation from that set generates a deviation from the critical state itself. In some ecological applications, however, it has been suggested that the critical state (i.e., the scale-free distribution) is robust to changes in the parameter state. That is, rather than a particular threshold at which the critical state occurs, there is a range of parameter states for which the criticality occurs. Such a state is referred to as "robust criticality" (Roy et al., 2003; Kefi et al., 2011; Pascual et al., 2002).

An example of this sort of dynamic structure is a general stochastic cellular automata model, inspired by our field work on an arboreal ant, *Azteca instabilis*, which nests in shade trees of a coffee plantation (Vandermeer et al., 2008). When instantiated with field data, the model appears to mimic both the spatial pattern and population density actually found in field surveys. That model serves to introduce the basic framework of various forms of self-organization and makes evident the relationship between spatial pattern and a catastrophic transition to a separate regime (extinction of the species), as we describe presently.

The model begins with a square lattice of size L, with lattice points {i, j}. Individual cells are affected by the nearest 8 neighbors (the Moore neighborhood), and the population density at a cell at time t is defined as the number of cells occupied in the Moore neighborhood of that cell. Thus, if B(i, j) is the binary variable indicating occupancy of cell i,j, the population density of the cell at i,j is given as

$$N_t(i,j) = \sum_{a=i-1}^{a=i+1} \sum_{b=i-1}^{b=i+1} B(a,b)$$

The probability that an empty cell at point (i, j) will become occupied is,
$$P_O = \alpha_o + \alpha N(i,j)$$

which in the end may be nonlinear or linear (as in the present case) and the probability that an occupied cell at point (i, j) will die is,

$$P_D = \gamma + \beta N(i,j),$$

where all four parameters are assumed to be positive real numbers. Density-independent occupations are stipulated by $\alpha_0$ which, by definition, is the measure of the "non-local" nature of the model (with $\alpha_0 = 0$ there are only local interactions). The cause of mortality is "epidemic" in the sense that it increases as the local population increases.

A variety of spatial situations in nature (in addition to the one in Vandermeer et al., 2008) exist that broadly correspond to this simple set of assumptions, as mentioned earlier. Any ecological system potentially could have its spatial structure described by this simple formula, a formula that is qualitatively similar to Turing's insight of the activator/repressor dialectic. Note that in this basic model the parameters that define local density-dependent effects are $\alpha$ and $\beta$, while the parameters representing global effects are $\gamma$, the global density-independent mortality rate and $\alpha_0$, the parameter that represents global mixing of the population.

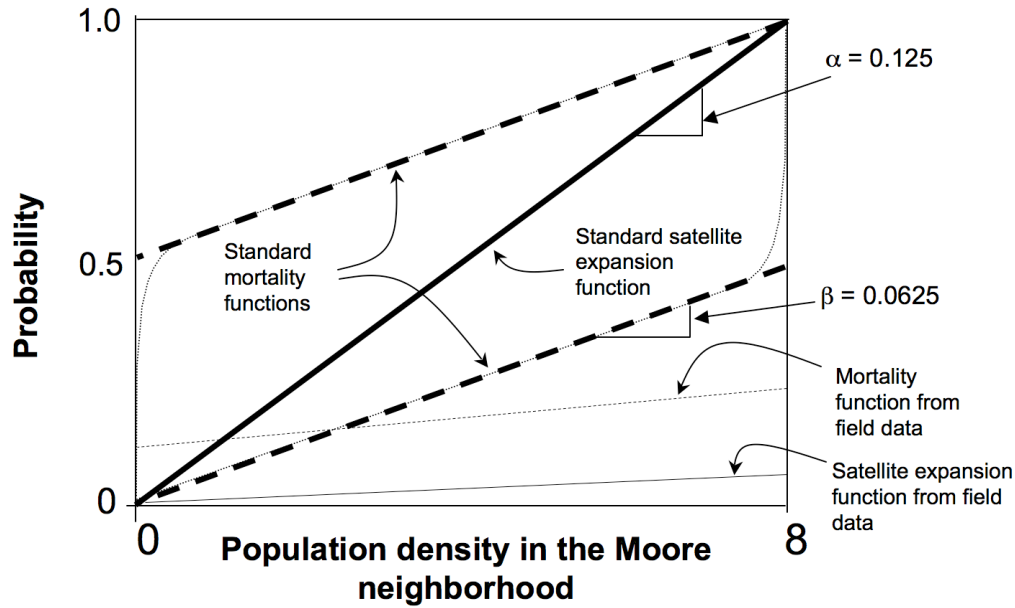

Figure 1. Basic parameter values for the standard model and for the field instantiated model (Vandermeer et al., 2008). Note that in the standard model $\alpha_0 = 0$, which is to say all interactions are completely local. The light dotted curves are illustrations of how a nonlinear model could approximate the linear forms used here.

Possible arrangements of the model can be most easily visualized on a graph of P versus N (fig 1). Since the range of N is 0 -- 8, two critical values for the parameters are obvious, as indicated in figure 1 as what we refer to as the "standard model," that is, $\alpha = 0.125$ (the value for which $P_0(8) = 1.0$) and $\beta = 0.0625$ (the value for which the range of $\gamma$s range from 0 to 0.5 and span all possible values in which the mortality and occupancy functions cross). Other values for $\alpha$ would effectively indicate a nonlinear satellite expansion function in that the intersection of the function at probability = 1.0 is at a population density less than 8, or the intersection of the function at population density of 8 is less than a probability of 1.0. The mortality functions in the standard model are effectively nonlinear in this sense (see dotted curves in figure 1), although they are incorporated in the model as the strict linear functions as illustrated in figure 1. Also included in figure 1 are the functions as computed from field data of the arboreal ant, *A. instabilis*, in a 45 ha plot in southern Mexico (Vandermeer et al., 2008).

We choose to measure spatial pattern as agreement with a power function. For many model parameters there is an evident deviation from that expectation, especially near the catastrophic transition point (as noted for other similar models, e.g., Kefi, et al., 2011), but also at parameter values far from that point. For comparative purposes, we employed the null model of Kefi et al., (2011), with non-local probabilities of occupying an empty site or extinguishing an occupied site.

The basic structure of this model system is summarized in figure 2. The null model (pictured in black on the base mortality rate graph) shows a simple decline in population density as the base mortality increases, not a surprising fact. Furthermore the null model generates patterns that are approximated fairly well by a power function, although given the basically random structure of the null model, the range of cluster sizes is too small to warrant the conclusion that the system is truly scale free (Newman, 2005; Clauset et al., 2009). Nevertheless, for the standard model, an almost perfect fit to the power function occurs at the point at which a spanning cluster emerges, as reported elsewhere (Kefi et al., 2011).

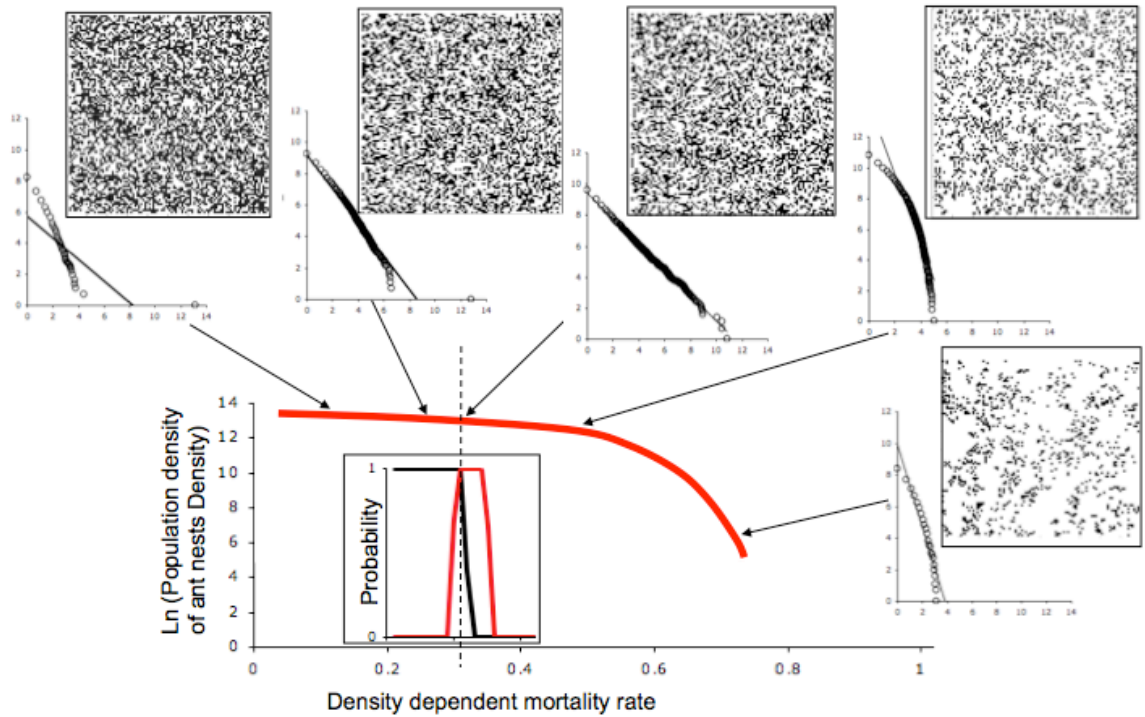

*Figure 2. Natural log of the population density as a function of mortality rate (γ), for both the null model (black curve) and the standard model (red curve), where the dashed curve indicates unstable points. Five examples from five different values of g along with spatial pattern for the standard model and two examples from two different values of g along with spatial pattern for the null model (frequency diagrams are cumulative). Note the single very large cluster in each of the two upper left graphs, to the left of the percolation point. Insert graph is probability of having a spanning cluster (black curve) and probability of the coefficient of determination for a power function fit being larger than 0.95 (red line), plotted on the same base mortality rate axis. Note that a deviation from the power law cannot be assured (i.e., a coefficient of determination less than .95) for a broad range of g which is the signal of "robust" criticality. Unstable set of points (red dashed curve) determined by 50% of simulations increasing at that value of g (i.e., points below the dashed curve descended to zero more than 50% of the time, points above the dashed curve increased to the equilibrium value more than 50% of the time).*

For the standard model this point is approximately γ = 0.33 (the top middle cluster size graph in figure 2).

The standard model follows the pattern evident in other similar spatial models (Pascual et al., 2002; Kefi et al., 2011; Rietkerk et al., 2004), with, as expected, a very high value of $R^2$ at γ = 0.33 (where the spanning cluster emerges). The population density drops catastrophically to zero at approximately

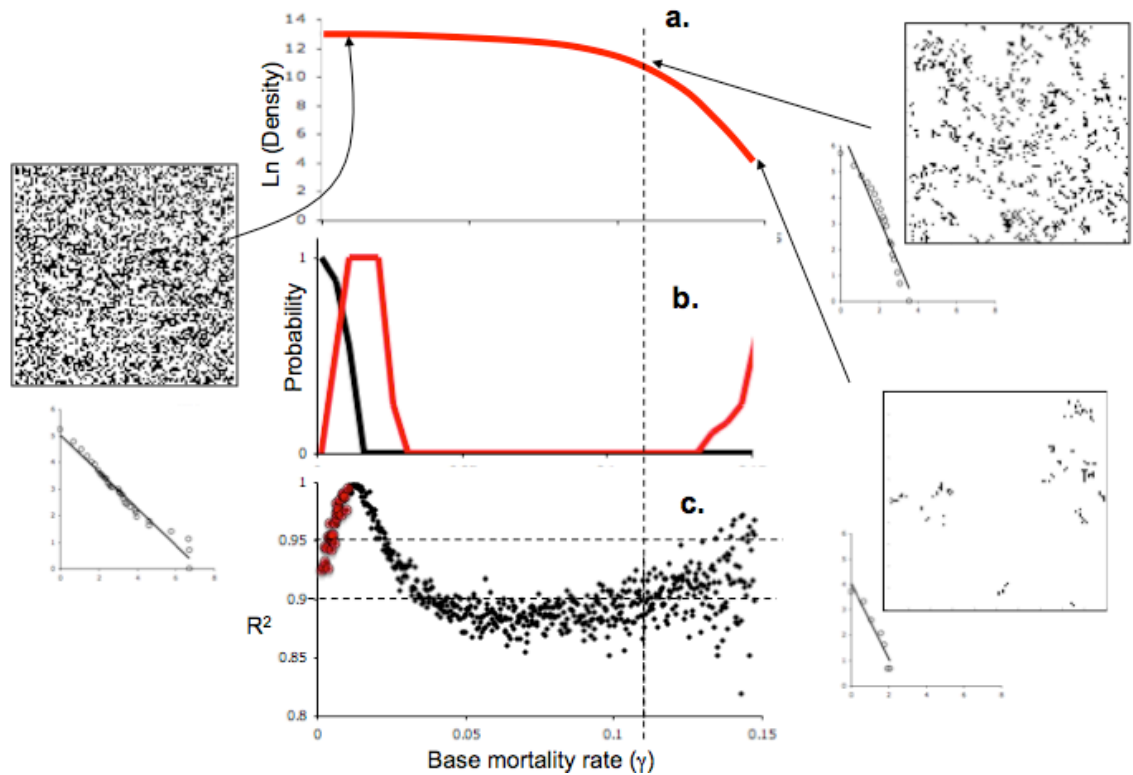

*Figure 3. Model with field instantiated parameters (except γ, the base mortality rate, which is used as a tuning parameter). a. Log of the population density as a function of mortality rate (g), with associated fits to the power function shown both graphically (with examples of the associated spatial patterns) and statistically with fits to the power function for a single example. b. Probability of a spanning cluster (black curve) and failure to reject the power law (i.e., $R^2$ > 0.95) red line. c. 600 separate simulations (500 iterations each) on a 1000 x 1000 lattice, black symbols where there was no spanning cluster and red symbols where there was at least one spanning cluster.).*

γ = 0.72. It is of further interest to note that the probability of $R^2$ > 0.95 remains at 1.0 for a range of values of γ, the measure of robust criticality (Kefi et al., 2011; Pascual et al., 2002).

Most importantly, failure to reject the power function hypothesis (as measured by $R^2$ > 0.95) has a regular form that decays as γ either increases or decreases. When densities are larger than the density that gives rise to a spanning cluster (i.e., with γ < 0.32), the spanning cluster itself begins connecting with other big clusters, generating a fat-tailed distribution, thus causing a deviation from the power law. When densities are smaller than the spanning cluster density, $R^2$ also decays, but the reason is because both the size of the largest clusters and the number of the smallest clusters are smaller than expected from a perfect scale-free distribution. Others (Kefi et al., 2007; 2011) have characterized this pattern as a truncated power function or a power function with an exponential tail, and have suggested that its appearance might be a sign of impending regime change. For this model (with the standard parameter set) we see that this pattern does not hold. Indeed, deviation from the power function begins very near the spanning cluster density and remains more or less consistent with increasing γ until the catastrophic collapse.

As indicated in figure 1, the parameters for this model as estimated from the field in Vandermeer et al. (2008) deviate significantly from what we refer to as the standard model. In figure 3 we illustrate the structure of the model using the field instantiated values of $\alpha_o$, $\alpha$, and $\beta$, using $\gamma$ as a tuning parameter (the field value of $\gamma$ is 0.116). For this parameter set it is evident that we are near the edge of the possible percolation, which is to say spanning cluster. Only when the base mortality rate is very small do we see the generation of spanning clusters, with the concomitant generation of cluster size distributions that appear to be scale free (judging from the coefficient of determination, an imperfect criterion, to be sure). Especially interesting is the pattern of individual determinations of the coefficients of determination (fig. 3c), where it appears that the percolation point is the precise point at which the scale free distribution emerges and the deviation away from a good fit to the power function occurs quite rapidly with either an increase or decrease in the base mortality rate. The conclusion one might draw from figure 3b, that the criticality is robust, is thus questionable.

Furthermore, the conclusion previously drawn that the field data and model with the field parameters represents robust criticality (Vandermeer et al., 2008) itself is questionable. If the robust region of criticality is near a base mortality rate of approximately 0.01 (see figure 3c), this must be compared to the field instantiated figure of 0.11, indicated with a vertical dashed line in figure 3c. Yet one can see how the conclusion of robust criticality might be erroneously drawn here if the critical $R^2$ is set at 0.9, wherein more than half of the simulations did indeed imply, probably incorrectly, a power function. Note that, in principle, the robust criticality argument is that a fit to a power function should be near the percolation point and that fit should extend beyond the precise parameter value that generates the percolation. Here we see a dramatic deviation from the power function fit as $\gamma$ increases even a small amount (fig. 3c).

**Discussion:**

We argue that the ant *Azteca instabilis*, forms spatial patterns based on a Turing-like dynamic (Vandermeer et al., 2008), which can be modeled with a simple stochastic cellular automata model. The CA model proposed to mimic the Turing-like dynamics is extraordinarily simple and consists of the tendency to move to neighboring cells when local density is high coupled with the tendency to die when local density is high. Thus, both "birth" and death are density dependent. In this article we present a systematic study of the behavior of this model, first establishing what we refer to as the standard model, a series of parameter ranges that make sense under the assumption of linearity (figure 1).

The overall behavior of the model corresponds to other, similar, models (figure 2) in most respects. Note, however, that our model is substantially simpler than many of the other published spatially-explicit models. Given that it generates behavior similar to many if not most of the spatially-explicit single-population models, we propose that its simplicity suggests that it may be considered as a baseline from which other more complicated models incorporating specific aspects of specific populations can be explored. At its foundation, we argue, the operation of the model is similar, at least metaphorically, to the basic Turing process where local activation is balanced by local repression, where activation here is local spatial expansion and repression is the action of some density-dependent mortality factor.

As expected, the model generates a percolation (spanning cluster) at a particular population density and at that percolation point the distribution of sizes of clusters of cells conforms quite closely to a power law. Decreases in that population density are accompanied by deviation from the power law form, corresponding to a truncated power law (or, a power law with an exponential tail), much as has been reported elsewhere (Kefi et al., 2011). The prospect of using that deviation as a signal of impending regime change has been discussed elsewhere (Kefi et al., 2007). It is also of interest to note that there is a completely distinct form of deviation from the power law as the mortality rate decreases and population consequently increases. There we see a pattern in which the spanning cluster continues expanding, effectively swallowing up other clusters and becoming much larger than would be expected from a power law. Concomitantly the number of isolated cells becomes proportionally smaller than what would be expected from a strict power law (figure 2).